\documentclass[prd,twocolumn,showpacs,preprintnumbers,amsmath,amssymb,superscriptaddress,floatfix,nofootinbib]{revtex4-2}
\usepackage{graphicx}
\usepackage{amsmath}
\usepackage{amsfonts}
\usepackage{amssymb}
\usepackage{color}
\usepackage{multirow}
\usepackage[colorlinks, citecolor=blue,anchorcolor=red,menucolor=red,linkcolor=red,filecolor=red,runcolor=red,urlcolor=blue,frenchlinks=red]{hyperref}

\usepackage{subfigure}

\begin{document}

\title{Role of the scalar $f_0(980)$ in the process $D_s^+ \to \pi^{+} \pi^{0} \pi^{0}$}

\author{Han Zhang}
\affiliation{School of Physics and Microelectronics, Zhengzhou University, Zhengzhou, Henan 450001, China}

\author{Yun-He Lyu}
\affiliation{School of Physics and Microelectronics, Zhengzhou University, Zhengzhou, Henan 450001, China}

\author{Li-Juan Liu}
\email{liulijuan@zzu.edu.cn}
\affiliation{School of Physics and Microelectronics, Zhengzhou University, Zhengzhou, Henan 450001, China}

\author{En Wang}
\email{wangen@zzu.edu.cn}
\affiliation{School of Physics and Microelectronics, Zhengzhou University, Zhengzhou, Henan 450001, China}

\begin{abstract}

Based on the BESIII measurements on the reaction of $D_s^+\to \pi^+\pi^0\pi^0$, we investigate this process by considering the $S$-wave pseudoscalar-pseudoscalar interaction within the unitary chiral  approach, and the contributions from the intermediate resonances $f_0(1370)$ and $f_2(1270)$. Our calculation could reasonably reproduce the experimental data, and our results imply that the $f_0(980)$, dynamically generated from the $S$-wave pseudoscalar-pseudoscalar interaction, plays an important role in this process, and the contributions from the intermediate resonances $f_0(1370)$ and $f_2(1270)$ are also necessary. The more precise measurements of this process in future could shed light on the nature of the $f_0(1370)$ and $f_2(1270)$.
\end{abstract}


\maketitle

\section{Introduction} \label{sec:Introduction}
In last decades, the constituent quark model successfully explains the composition of the most mesons~\cite{Workman:2022ynf}. However, the properties of the light scalar mesons are difficult to be described within the constituent quark model. For the scalar mesons with masses less than 1~GeV, $f_0(500)$, $a_0(980)$, $f_0(980)$, and $K(700)$, there are  different explanations, such as the $q\bar{q}$ states, tetraquark states, or molecular states~\cite{tHooft:2008rus,Close:2002zu,Wang:2017pxm,Deng:2012wj}. In addition, the nature of the isospin-zero scalar mesons with larger masses, $f_0(1370)$, $f_0(1500)$, and $f_0(1710)$, are also difficult to be fully understood~\cite{Workman:2022ynf}. Recently, the BESIII and Belle/Belle II have accumulated lots of experimental information about the charmed hadrons decays, which provides an important platform to explore the internal structures of those scalar mesons~\cite{Wang:2022nac,Wang:2021naf,Duan:2020vye,Wang:2020pem}. 

For instance, the BESIII Collaboration have observed the scalar $S(1710)$\footnote{It should be stressed that BESIII does not distinguish between the $a_0(1710)$ and $f_0(1710)$ in the $D^+_s\to K^0_SK^0_S\pi^+$, and denotes the combined state as $S(1710)$~\cite{BESIII:2021anf}.} with the mass of $1723\pm 11\pm2$~MeV and width of $140\pm 14\pm 4$~MeV in the process of $D_s^+\to K^0_S K^0_S \pi^+$~\cite{BESIII:2021anf}. Later, the similar structure was also observed in the process of $D_s^+\to K^+ K^0_S \pi^0$ by the BESIII Collaboration, associated with a new scalar state $a_0(1817)$ with the mass of $1817 \pm 8\pm 20$~MeV and width of $97\pm 22 \pm 15$~MeV~\cite{BESIII:2022npc}. The $\sim100$~MeV mass discrepancy between $S(1710)$ and $a_0(1817)$ causes the doubt whether they are isospin partners. However, the line-shapes of the $a_0(1817)$ and $S(1710)$ in the $K^+K^0_S$ and $K^0_SK^0_S$ invariant mass distributions are so similar that the measured mass discrepancy of the $a_0(1817)$ and $S(1710)$ may be due to their peaks' being very close to the threshold of the $K\bar{K}$ mass spectrum. We have investigated both reactions, and found that the BESIII measurements of both reactions could be well reproduced by regarding the intermediate state $a_0(1817)$ as the $K^*\bar{K}^*$ molecular state~\cite{Zhu:2022duu,Zhu:2022wzk}, which is also supported by the study of Ref.~\cite{Dai:2021owu}.

The BESIII Collaboration recently performed an amplitude analysis of the process $D_s^+ \to \pi^{+} \pi^{0} \pi^{0}$, and determined the branching fraction $\mathcal{B}$($D_s^+ \to \pi^{+} \pi^{0} \pi^{0}$) = (2.8 $\pm$ 0.4 $\pm$ 0.4)$\%$~\cite{BESIII:2021eru}, which improves the precision with a factor of two compared to the results of the CLEO Collaboration~\cite{CLEO:2009vke}. The significant $f_{0}(980)$ signal and an enhancement structure around 1300~MeV were observed in the $\pi^{0}\pi^{0}$ invariant mass spectrum, and the later structure could be associated with the intermediate resonances $f_0(1370)$ and $f_2(1270)$. The scalar $f_0(980)$ could be explained as the $K\bar{K}$ molecular state dynamically generated from the $S$-wave pseudoscalar-pseudoscalar interaction~\cite{Oller:1997ti,Oller:1998hw}, which was supported by many works~\cite{Wang:2022nac,Wang:2021naf,Duan:2020vye,Wang:2020pem}. In addition, the resonance $f_0(1370)$ is a broad state, and its mass and width are not well established. There are various explanations for the nature of $f_0(1370)$, such as the $n\bar{n}$ state, molecular state~\cite{Molina:2008jw,Geng:2008gx,Garcia-Recio:2013uva},  $s\bar{s}$~\cite{Celenza:2000uk}, and glueball~\cite{Janowski:2014ppa,Minkowski:1998mf}. In Refs.~\cite{Molina:2008jw,Geng:2008gx,Garcia-Recio:2013uva}, $f_2(1270)$ is also explained as the bound system  of two vector mesons, which is challenged in Ref.~\cite{Gulmez:2016scm}.
Therefore, the study of the process $D_s^+ \to \pi^{+} \pi^{0} \pi^{0}$ should be helpful to deepen our understanding about the nature of the $f_0(980)$, $f_0(1370)$, and $f_2(1270)$.

In this work, we will investigate the process $D_s^+ \to \pi^{+} \pi^{0}\pi^{0}$ by considering the scalar $f_0(980)$ generated from the $S$-wave pseudoscalar-pseudoscalar interaction within the chiral unitary approach and the contributions from the intermediate resonances $f_0(1370)$ and $f_2(1270)$. 

This paper is organized as follows. In Sec.~\ref{sec:Formalism}, we will present the mechanism for the process $D_s^+ \to \pi^{+} \pi^{0}\pi^{0}$, and results and discussions will be given in Sec.~\ref{sec:Results}, followed by a short summary in the last section.

\section{Formalism}  \label{sec:Formalism}

\begin{figure}[tbhp]
\begin{center}
\includegraphics[scale=0.7]{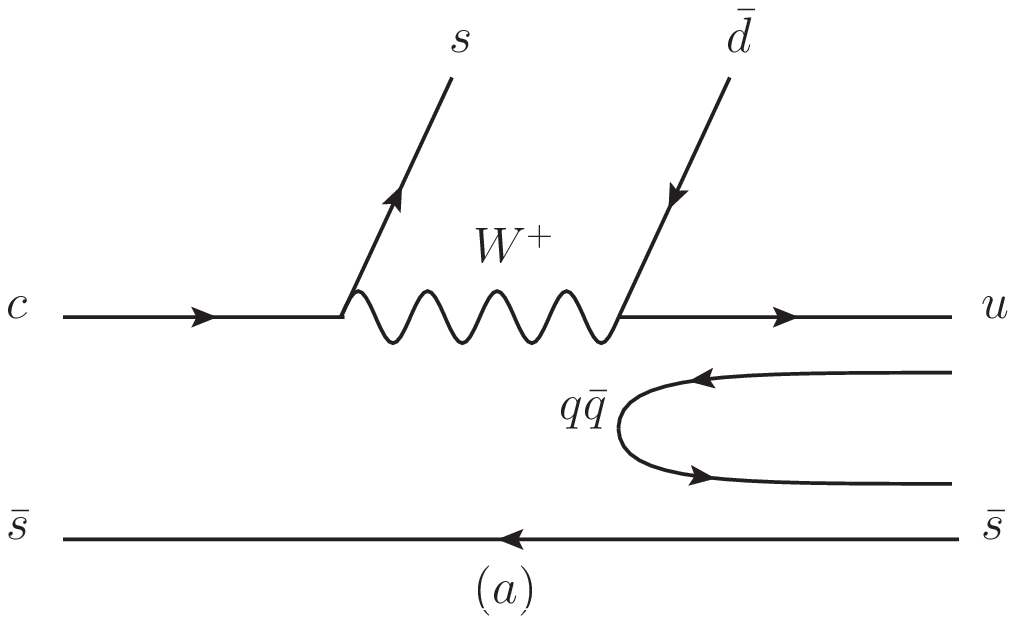}{\label{a}}
\includegraphics[scale=0.7]{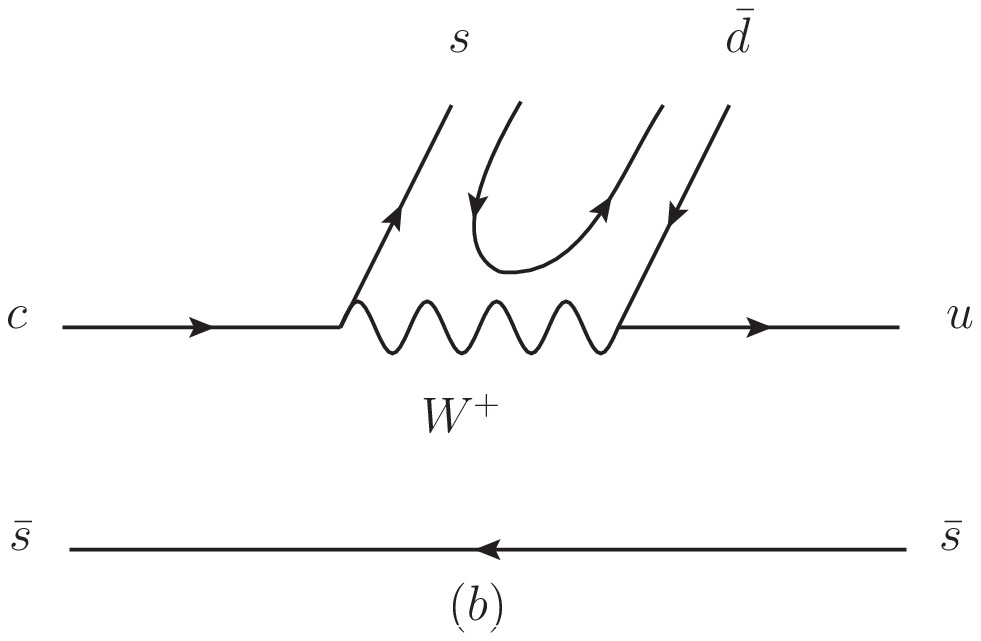}{\label{b}}
\includegraphics[scale=0.7]{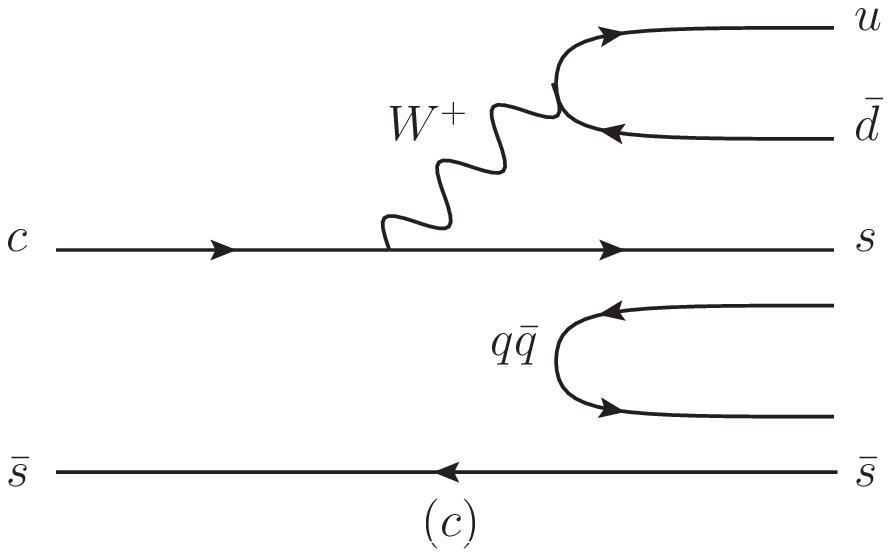}{\label{c}}
\includegraphics[scale=0.7]{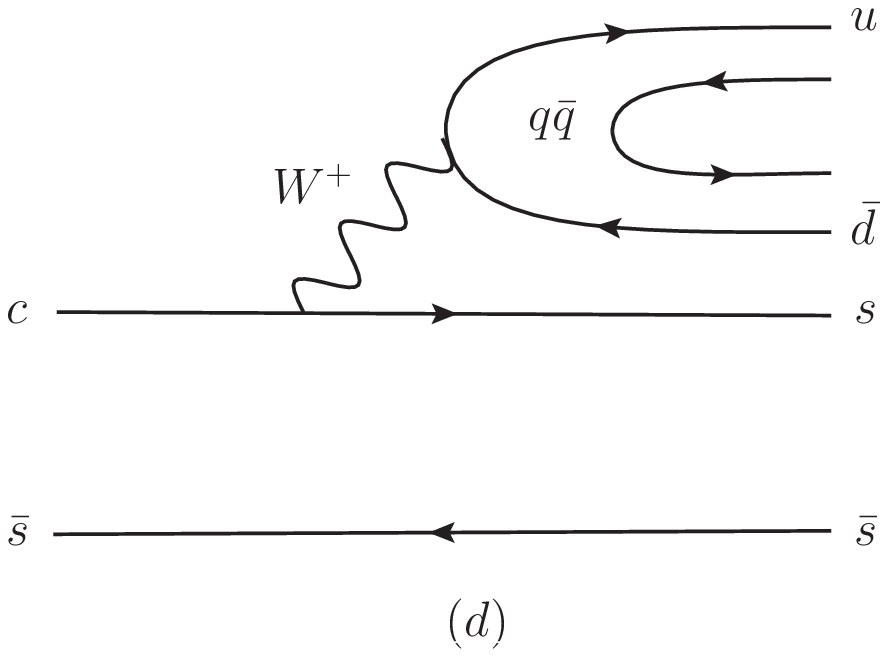}{\label{d}}
\end{center}
\caption{The quark level diagrams for the $D_{s}^{+}$ weak decays. (a) The internal emission of $D_{s}^{+} \to \bar{K}^{0}u\bar{s}$ and hadronization of the $u\bar{s}$ through $q\bar{q}$ with vacuum quantum numbers. (b) The internal emission of $D_{s}^{+} \to K^{+}s\bar{d}$ and hadronization of the $s\bar{d}$ through $q\bar{q}$  with vacuum quantum numbers. (c) The external emission of $D_{s}^{+} \to \pi^{+}s\bar{s}$ and hadronization of the $s\bar{s}$ through $q\bar{q}$ with vacuum quantum numbers. (d) The external emission of $D_{s}^{+} \to u\bar{d}\eta$ and hadronization of the $u\bar{d}$ through $q\bar{q}$ with vacuum quantum numbers.}
\label{Fig:QuarkLevel}
\end{figure}

In this section, we will present the mechanism of the process $D_s^+\to \pi^+\pi^0\pi^0$. This process will happen via three steps, the weakly decay, the hadronization, and the final state interactions~\cite{Wang:2021naf,Duan:2020vye,Zhu:2022duu,Zhu:2022wzk}. Firstly, the $c$ quark of the initial $D_s^+$ weakly decays into an $s$ quark and a $W^+$ boson, and subsequently the $W^+$ boson decays into a $\bar d u $ quark pair.  Then all the quarks, together with the quark pair $q\bar q $ (=$u\bar u$ + $d\bar d$ + $s\bar s$ ) created from the vacuum with the quantum numbers $J^{PC}=0^{++}$, hadronize into the hadrons, which could be classified as the $W^+$ internal emission of Figs.~\ref{Fig:QuarkLevel}(a-b) and the $W^+$ external emission of Fig.~\ref{Fig:QuarkLevel}(c-d).

For the $W^+$ internal emissions of Figs.~\ref{Fig:QuarkLevel}(a) and~\ref{Fig:QuarkLevel}(b), the $s\bar{d}$ or $u\bar{s}$ hadronize into the $\bar{K}^0$ or $K^+$, and the hadronization of the other quarks could be expressed as
	\begin{eqnarray}
	&&\sum_{i}u(\bar q_{i} q_i)\bar s=\sum_{i}M_{1i}M_{i3}=(M^2)_{13},  \label{1} \\		
	&&\sum_{i}s(\bar q_{i} q_{i})\bar d=\sum_{i}M_{3i}M_{i2}=(M^2)_{32},	\label{2}
    \end{eqnarray}
where  $i=1,~2,~3$ correspond to the $u$, $d$, and $s$ quarks, respectively, and the $M$ is the $q\bar{q}$ matrix,
\begin{align}
	M=\left(\begin{array}{ccc}
		u\bar{u} & u\bar{d} & u\bar{s}\\
		d\bar{u} & d\bar{d} & d\bar{s} \\
		s\bar{u} & s\bar{d} & s\bar{s}
	\end{array}
	\right)\,.
	\label{eq:Mmatrix}
\end{align}
Within the $SU(3)$ flavor symmetry, the matrix $M$ can be written in terms of pseudoscalar mesons as~\cite{Duan:2020vye}
\begin{center}
	\begin{align}
		M=
		\left(\begin{array}{ccc}
			\frac{\pi^0}{\sqrt{2}} +\frac{\eta}{\sqrt{3}}+\frac{\eta^\prime}{\sqrt{6}}& \pi^+ & K^+\\
			\pi^-& -\frac{\pi^0}{\sqrt{2}} +\frac{\eta}{\sqrt{3}}+\frac{\eta^\prime}{\sqrt{6}} & K^0\\
			K^-& \bar{K}^0 & - \frac{\eta}{\sqrt{3}}+\frac{2\eta^\prime}{\sqrt{6}}
		\end{array}
		\right)\,.
		\label{Mmatrix}
	\end{align}
\end{center}

Since the $\eta^\prime$ has a large mass and does not play a role in the generation of $f_0(980)$, we ignore the $\eta^\prime$ component in this work. Then the Eqs.~(\ref{1}-\ref{2}) can be rewritten as,
 \begin{eqnarray}
	&&(M^{2})_{13}=\frac{1}{\sqrt{2}}K^+\pi^0 + \pi^+K^0,  \\
	&& (M^{2})_{32}=K^-\pi^+ - \frac{1}{\sqrt{2}}\bar{K^0}\pi^0. 
  \end{eqnarray}
 
For the $W^+$ external emission of Fig.~\ref{Fig:QuarkLevel}(c), the quark $u\bar{d}$ of the $W^+$ decay hadroinze into $\pi^+$, and the $s\bar{s}$ pair, together with the created $q\bar{q}$ pair, hadronize into the states as, 
 \begin{eqnarray}
	\sum_{i}s(\bar q_{i} q_i)\bar s &=&\sum_{i}M_{3i}M_{i3}=(M^2)_{33}  \nonumber \\		
	&=& K^+K^-+K^0\bar{K}^0+\frac{1}{3}\eta \eta.
    \end{eqnarray}
    
    For the $W^+$ external emission of Fig.~\ref{Fig:QuarkLevel}(d), the $s\bar{s}$ pair could hadronize into the $\eta$ meson, and the quark $u\bar{d}$ of the $W^+$ decay, together with the created $q\bar{q}$ pair, hadroinze into $\pi^+\eta$, which contributes to the process $D_s^+\to \pi^+\eta\eta$. Thus, we have,, 
 \begin{eqnarray}
	\sum_{i}u(\bar q_{i} q_i)\bar d (s\bar{s}) &=&\sum_{i}M_{1i}M_{i2} M_{33} \nonumber \\		
	&=&(M^2)_{12} \left(-\frac{1}{\sqrt3}\eta\right)= - \frac{2}{3}\pi^+\eta \eta.
    \end{eqnarray}
  
\begin{figure}[tbhp]
\begin{center}
\includegraphics[scale=0.7]{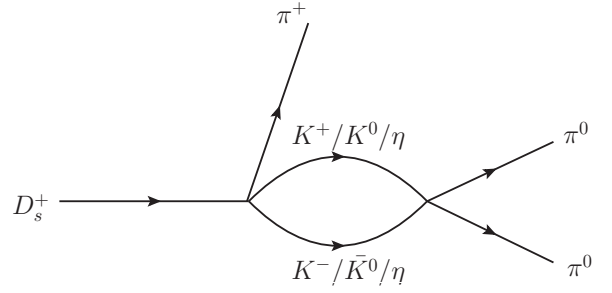}
\end{center}
\caption{The S-wave final state interaction of the meson pairs. } \label{Fig:loop}
\end{figure}

Then the processes of the $D^+_s$ decaying into all possible states could be expressed as
\begin{eqnarray}
	H^{(a)} &=&V_{cs}V_{ud}\left(\frac{1}{\sqrt{2}}\pi^0K^{+}+\pi^{+}K^{0} \right)\bar{K}^0, \\
    H^{(b)} &=&V_{cs}V_{ud}\left(\pi^{+}K^{-}-\frac{1}{\sqrt{2}}\pi^0\bar {K}^0 \right) K^+, \\
    H^{(c)}&=& CV_{cs}V_{ud}\left(K^+K^-+K^0\bar{K^0}+\frac{1}{3}\eta \eta \right)\pi^+, \\
      H^{(d)}&=& CV_{cs}V_{ud}\left(-\frac{2}{3}\eta \eta \right)\pi^+,
\end{eqnarray}
where $V_{ud}$ and $V_{cs}$ are the  CKM matrix elements.
Since the external emission of $W^+$ bosons is color-favored relative to the $W^+$ internal emission, an extra color factor $C$ can be introduced to account for the relative weight of the $W^+$ external emission with respect to the $W^+$ internal emission. For the $W^+$ external emission, the $u\bar d$ quark pair from the $W^+$ decay can form the color singlet $\pi^+$, and the $u$ and $\bar{d}$ have three choices of colors, while for the $W^+$ internal emission, the $u$, $\bar{d}$, $s$ and $\bar s$ quarks from the $W^+$ decay have fixed colors. Thus, the factor $C$ is taken to be 3 in this work~\cite{Wei:2021usz,Dai:2018tgo,Zhang:2020rqr,Dai:2018nmw}. Now we have all the possible components after the hadronization,\footnote{Here we neglect the components $\pi^0\bar{K}^0 K^+$. Indeed,  the interaction of the $\bar{K}^0 K^+ $ to $\pi^+\pi^0$ can be given by  the intermediate resonances, which is shown very small by BESIII~\cite{BESIII:2021eru}.} 
\begin{align}
	H=& H^{(a)}+H^{(b)}+H^{(c)}+H^{(d)}\nonumber\\
	=&V_{p}V_{cs}V_{ud}\left\{C( K^+K^-+K^0\bar{K}^0+\frac{1}{3}\eta\eta )\pi^+ \right.\nonumber \\ 
	&\left. + \bar{K^0}K^0\pi^+ + K^+ K^-\pi^+\right\} \nonumber \\ 
	=&V_{p}V_{cs}V_{ud}\left\{
	(C+1)\left( K^+K^-\pi^++K^0\bar{K}^0\pi^+ \right) \right.\nonumber \\ 
	&\left. -\frac{C}{3}\eta\eta \pi^+ \right\},
\end{align}
where the $V_p$ is the factors of the production vertices containing all the dynamics. After the preliminary weakly decay, the meson pairs of $K^{+}K^{-}$, $K^{0}\bar{K}^0$, and $\eta\eta$ could undergo the $S$-wave final state interaction to give rise to the $\pi^{0}\pi^{0}$ final state, where the  scalar meson $f_0(980)$ could be dynamically generated as shown in Fig.~\ref{Fig:loop}.

The amplitude of the $S$-wave pseudoscalar-pseudoscalar interaction, generating the scalar $f_0(980)$, now can be written as,
\begin{eqnarray}
	{\mathcal{M}}_{f_{0}(980)}&=&V_p\left[(C+1) G_{K^+K^-}t_{K^+K^-\to \pi^0\pi^0}\right.\nonumber \\
	&& +(C+1)  G_{K^0\bar{K}^0} t_{K^0\bar{K}^0 \to \pi^0\pi^0 }\nonumber \\
	&&\left. -\frac{C}{3} G_{\eta\eta} t_{\eta\eta \to \pi^0\pi^0 } \right] , \label{eq:fullamp}
\end{eqnarray}
where $G_i$ is the loop function of two-meson propagator, and $t_{i\to j}$ is the transition amplitude of the $i$-channel to $j$-channel, both of which are the functions of the $\pi^0 \pi^0$ invariant mass $M_{\pi^0\pi^0}$. The loop function is given by,
\begin{equation}
	G_{i}= i  \int \frac{d^4 q}{(2 \pi)^4} \,
	\frac{1}{(P-q)^2 - m_1^2 + i \epsilon} \,
	\frac{1}{q^2 - m^2_2 + i \epsilon},\label{eq:loop}
\end{equation}
where $m_1$ and $m_{2}$ are the masses of the two mesons in the loop of the $i$-channel, and $P$ and $q$ are the four-momenta of the two-meson system and the second meson, respectively. The Mandelstam invariant $s=P^2=M_{\pi^0\pi^0}^2$.
The loop function of Eq.~(\ref{eq:loop}) is logarithmically divergent, and there are two methods to solve this singular integral, either using the three-momentum cut-off method, or the dimensional regularization method. The choice of a particular regularization scheme does not, of course, affect our argumentation.  In this work, we performed the integral for $q$ in Eq.~(\ref{eq:loop}) with a cut-off $|\vec{q}_{\rm max}|= 600$ MeV~\cite{Dias:2016gou,Liang:2014tia}. The transition amplitude $t_{i\to j}$ can be obtained by solving the Bethe-Salpeter equation in coupled channels,
\begin{equation}\label{BS}
	T=[1-VG]^{-1}V,
\end{equation}
where $V$ is a $5\times5$ matrix of the interaction kernel, we take five channels $\pi^+\pi^-$, $\pi^0\pi^0$, $K^+K^-$, $K^0\bar{K}^0$, and $\eta\eta$. The explicit expressions of the $5\times5$ matrix elements in $S$-wave are given by~\cite{Gamermann:2006nm,Liang:2014tia,Dias:2016gou,Ahmed:2020qkv}
\begin{eqnarray}
	&& V_{11}=-\frac{1}{2f^2}s,~~~V_{12}=-\frac{1}{\sqrt{2}f^2}(s-m^2_\pi), \nonumber \\
	&& V_{13}=-\frac{1}{4f^2}s,~~ V_{14}=-\frac{1}{4f^2}s,  \nonumber \\
	&&V_{15}=-\frac{1}{3\sqrt{2}f^2}m^2_\pi,~~~V_{22}=-\frac{1}{2f^2}m^2_\pi,\nonumber \\
	&& V_{23}=-\frac{1}{4\sqrt{2} f^2}s,~~~V_{24}=-\frac{1}{4\sqrt{2} f^2}s, \nonumber \\
	&& V_{25}=-\frac{1}{6f^2}m^2_\pi,~~ V_{33}=-\frac{1}{2f^2}s,~~~V_{34}=-\frac{1}{4f^2}s,\nonumber \\
	&&V_{35}=-\frac{1}{12\sqrt{2}f^2}(9s-6m^2_\eta-2m^2_\pi),\nonumber \\
	&& V_{44}=-\frac{1}{2f^2}s,~~V_{45}=-\frac{1}{12\sqrt{2}f^2}(9s-6m^2_\eta-2m^2_\pi),\nonumber \\
	&&V_{55}=-\frac{1}{18f^2}(16m^2_K-7m^2_\pi),
\end{eqnarray}
where $f=93$~MeV is the pion decay constant, and $s$ is invariant mass square of the meson-meson system, $m_\pi$, $m_K$, and $m_\eta$ are the masses of the pion, kaon, and $\eta$ mesons, respectively~\cite{Workman:2022ynf}. The unitary normalization $|\eta\eta> \to \frac{1}{\sqrt{2}}|\eta\eta>$ and $|\pi^0\pi^0>\to \frac{1}{\sqrt{2}}|\pi^0\pi^0>$ has been taken easily account for the identify of the particle when using the loop function $G$ without an extra factor~\cite{Liang:2014tia}.

In addition to the scalar $f_0(980)$, the BESIII has observed the enhancement structure around 1300~MeV in the $\pi^0\pi^0$ invariant mass distribution, which could be associated with the resonances $f_0(1370)$ and $f_2(1270)$. Hence, we will also take into account the contributions from the intermediate resonances $f_0(1370)$ and $f_2(1270)$.  

For the contribution of $f_0(1370)$ in the decay of $D_s^+ \to \pi^+\pi^0\pi^0$, we describe it using the Breit-Winger form,
\begin{eqnarray}
	\mathcal{M}_{f_0(1370)}=\frac{\alpha \times m^2_{f_0(1370)}}{m^{2}_{\pi^{0}\pi^{0}}-M^{2}_{f_{0}(1370)}+i\Gamma_{f_{0}(1370)}M_{f_{0}(1370)}} , \nonumber \\
	\label{eq:1370}
\end{eqnarray}
where $\alpha$ is the strength of the $f_0(1370)$. Considering that the mass and width of $f_0(1370)$ have large uncertainties~\cite{Workman:2022ynf}, we fix its mass to be 1300~MeV, the center position of the enhancement structure in the $\pi^0\pi^0$ invariant mass distribution of BESIII measurements~\cite{BESIII:2021eru}, and take the width as a free parameter.

Taking into account that the $f_2(1270)$ couples to the $\pi^0\pi^0$ in $D$-wave, we can write the contribution of this resonance as follows~\cite{Wang:2015pcn},
\begin{eqnarray}
	\mathcal{M}_{f_2(1270)}=\frac{\beta \times(3{\rm cos}^2\theta-1)\times{\tilde{p}_{\pi^0}^2}}{M^2_{\pi^0\pi^0}-M^2_{f_2(1270)}+i {M_{f_2(1270)}}\Gamma_{f_{2}(1270)}}, \nonumber \\
	\label{eq:termD}
\end{eqnarray}
where $\beta$ stands for the strength of the $D$-wave amplitudes. The mass and width of the $f_2(1270)$ are $M_{f_2(1270)}=1275.5$~MeV and $\Gamma_{f_2(1270)}=186.7$~GeV, taken from the Review of Particle Physics~\cite{Workman:2022ynf}. $\tilde{p}_{\pi^0}$ is the momentum of $\pi^0$ in the $\pi^0\pi^0$ rest frame, 
\begin{eqnarray}
	\tilde{p}_{\pi^0}=\frac{\lambda ^{1/2}\left(M^2_{\pi^0\pi^0},m^2_{\pi^0},M^2_{\pi^0}\right)}{2M_{\pi^0\pi^0}}. 
	\label{eq:a}
\end{eqnarray}
The parameter $\theta$ is the angle between the momentum of $\pi^0$ and $\pi^+$ in the rest frame of the $\pi^0\pi^0$ system~\cite{Wang:2015pcn}, 
\begin{eqnarray}	
	{\rm cos}\theta=\frac{M^2_{\pi^+\pi^0}-M^2_{D^+_s}-M^2_{\pi^0}+2\tilde p^0_{D^+_s}\tilde p^0_{\pi^0}}{2\tilde p_{\pi^{+}}\tilde p_{\pi^0}}, 
	\label{eq:b}
\end{eqnarray}
where $\tilde{p}^0_{D_s^+}$ ($\tilde{p}^0_{\pi^0}$) is the energy of $D_s^+$ ($\pi^0$) in the $\pi^0\pi^0$ rest frame, and $\tilde{p}_{\pi^+}=\tilde{p}_{D_s^+}$ is the $\pi^+$ ($D_s^+$) momentum in this same frame. We give the explicit forms for those variables below,
\begin{eqnarray}
	\tilde{p}_{\pi^+}&=& \tilde{p}_{D^+_s} = \frac{\lambda^{1/2}\left(M^2_{D^+_s}, M^2_{\pi^0\pi^0}, m^2_{\pi^+} \right)}{2M_{\pi^0\pi^0}} , \nonumber \\
	\tilde{p}^0_{D^+_s} &=& \sqrt{M^2_{D^+_s} + \tilde{p}^2_{D^+_s}}, \nonumber \\
	\tilde{p}^0_{\pi^0} &=& \frac{M_{\pi^0\pi^0}}{2},
	\label{eq:c}
\end{eqnarray}
where $\lambda(x,y,z)=x^2+y^2+z^2-2xy-2yz-2xz$.

Consequently, the total amplitude of $D_s^+ \to \pi^+\pi^0\pi^0$ can be described by
\begin{eqnarray}
	{\mathcal{M}}&=&a_{\rm bg}+{\mathcal{M}}_{f_0(980)}+{\mathcal{M}}_{f_0(1370)}+{\mathcal{M}}_{f_2(1270)},
	\label{eq:M}
\end{eqnarray}
where the constant $a_{\rm bg}$ is the background contribution\footnote{In Ref.~\cite{BESIII:2021eru}, BESIII has considered the background contributions from the $D_s^+\to \pi^+\pi^0 \eta $ where $\pi^0\eta$ is misreconstructed as $\pi^0\pi^0$. In addition, there are also the background from the non-resonant contributions.}. As a result, the amplitude of Eq.~(\ref{eq:M}) depends on the two independent invariant masses $M_{\pi^0\pi^0}$ and $M_{\pi^+\pi^0}$, the double differential width for the process $D_s^+ \to \pi^+\pi^0\pi^0$ is given by,
\begin{equation}
	\frac{d^2\Gamma}{dM_{\pi^0\pi^0}dM_{\pi^+\pi^0}}= \frac{M_{\pi^0\pi^0}M_{\pi^+\pi^0}}{(2\pi)^{3}8m^3_{D^+_{s}}}|\mathcal{M}|^2.
	\label{eq:Gamma^2}
\end{equation}
One can obtain the ${d\Gamma}/{dM_{\pi^0\pi^0}}$ and ${d\Gamma}/{dM_{\pi^+\pi^0}}$ by integrating the Eq.~(\ref{eq:Gamma^2}) over the other invariant mass variable with relations as follows,
 \begin{equation}
	\frac{d\Gamma}{dM_{\pi^0\pi^0}}=\int\frac{M_{\pi^0\pi^0}M_{\pi^+\pi^0}}{(2\pi)^{3}8m^3_{D^+_{s}}}|\mathcal{M}|^2dM_{\pi^+\pi^0}.	
  \label{eq:Gamma}	
 \end{equation}
 
With a given $M_{\pi^0\pi^0}$, the upper and lower bounds of the $M_{\pi^+\pi^0}$ are,
\begin{eqnarray}
   (M^2_{\pi^+\pi^0})_{\rm max}&=&\left(E^*_{\pi^+}+E^*_{\pi^0}\right)^2- \nonumber \\
   && \left(\sqrt{E^{*2}_{\pi^+}-m^{*2}_{\pi^+}}-\sqrt{E^{*2}_{\pi^0}-m^{*2}_{\pi^0}}\right)^2, \\
   (M^2_{\pi^+\pi^0})_{\rm min}&=&\left(E^*_{\pi^+}+E^*_{\pi^0}\right)^2- \nonumber \\
   && \left(\sqrt{E^{*2}_{\pi^+}-m^{*2}_{\pi^+}}+\sqrt{E^{*2}_{\pi^0}-m^{*2}_{\pi^0}}\right)^2,
	\label{eq:M^2}
\end{eqnarray}
here $E_{\pi^{+}}^{*}$ and $E^{*}_{\pi^{0}}$ are the energies of $\pi^{+}$ and $\pi^{0}$ in the $\pi^{0}\pi^{0}$ rest frame, respectively,
 \begin{eqnarray}
	E^{*}_{\pi^{+}}&=&\frac{m^{2}_{D^{+}_{s}}-M^{2}_{\pi^0\pi^0}-m^2_{\pi^+}}{2M_{\pi^0\pi^0}},  \nonumber \\
	E^{*}_{\pi^{0}}&=&\frac{M^{2}_{\pi^{0}\pi^0}-m^{2}_{\pi^0}+m^2_{\pi^0}}{2M_{\pi^0\pi^0}}.
	\label{eq:Epi}
\end{eqnarray}
Similarly, we can obtain the $\pi^{+}\pi^{0}$ invariant mass distribution.

\section{Numerical results and discussion}   \label{sec:Results}

\begin{figure}[htpb]
  	\begin{center}
  		\includegraphics[scale=0.65]{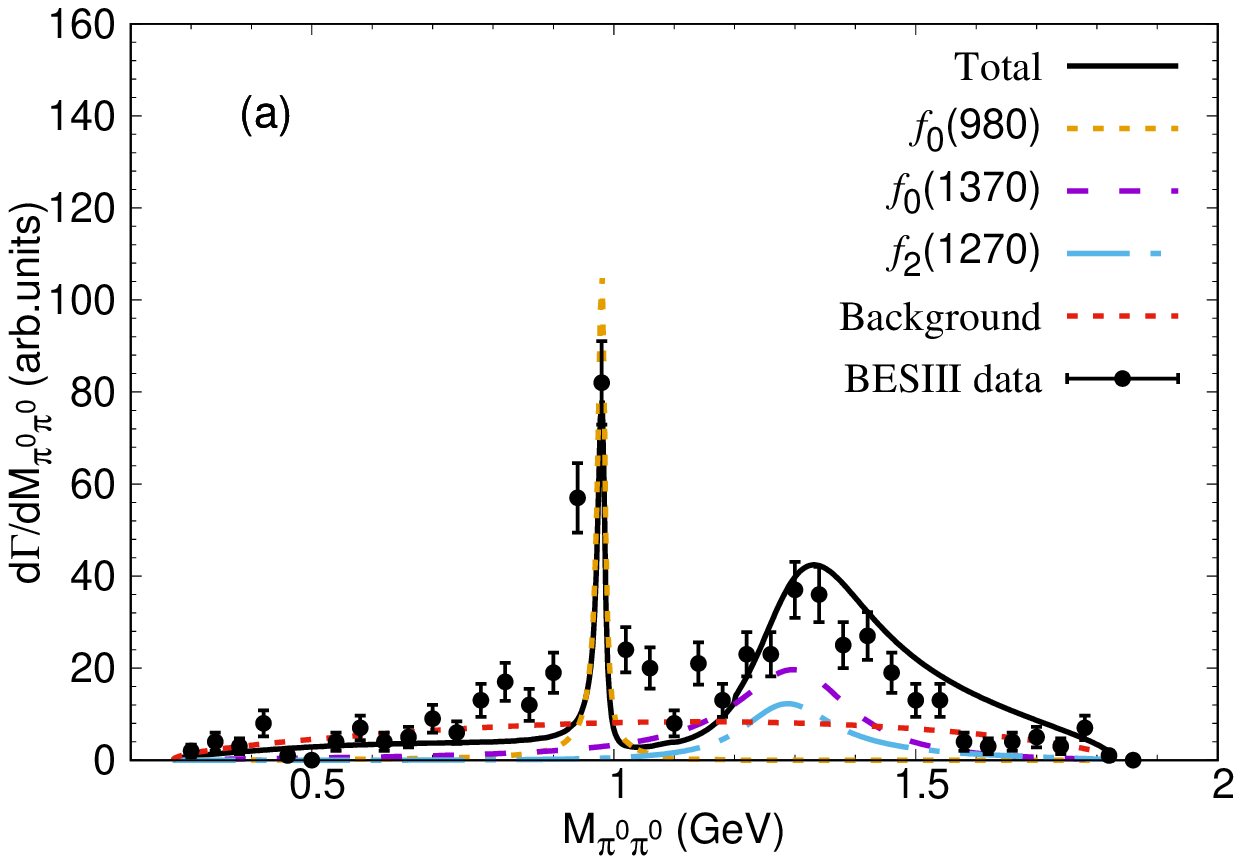}
  		\includegraphics[scale=0.65]{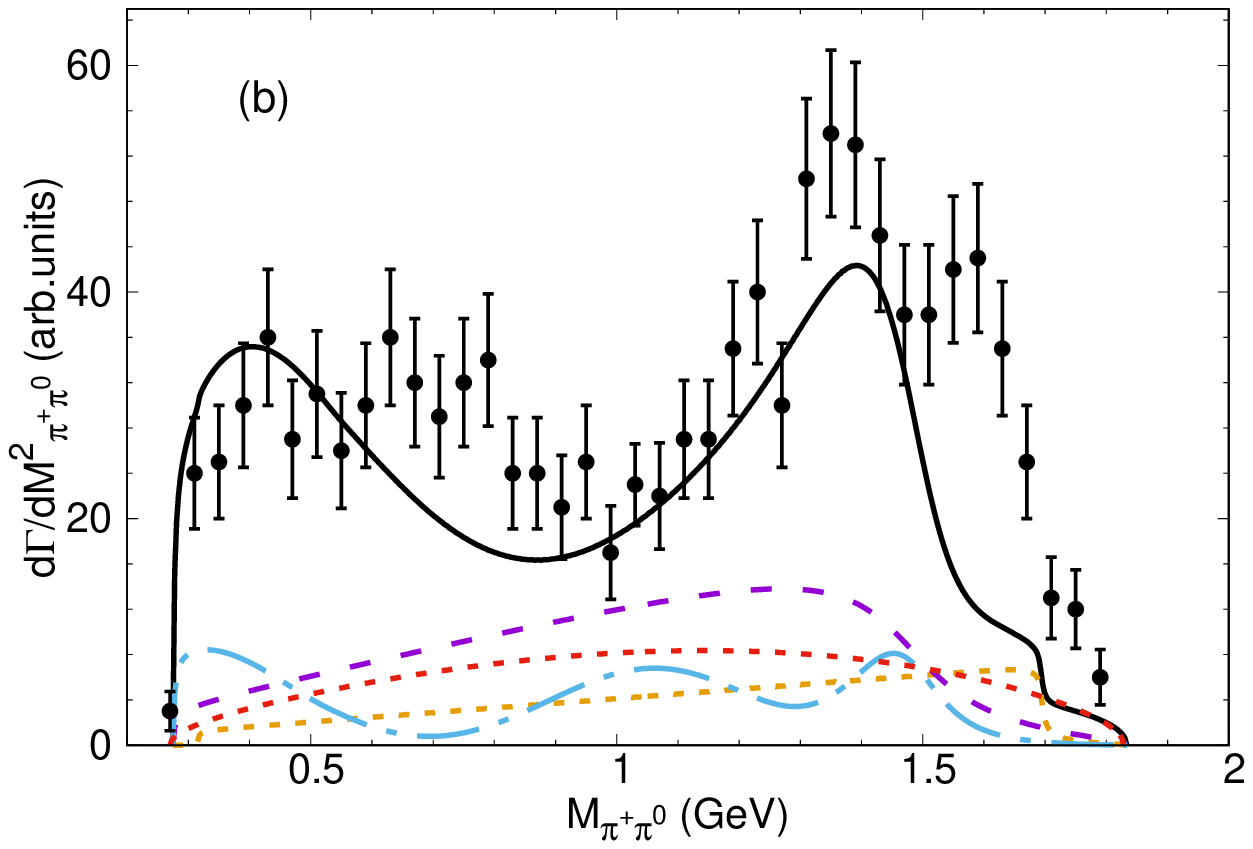}
  	\end{center}
  	\caption{\small {The $\pi^0\pi^0$ (a) and $\pi^+\pi^0$ (b) mass distributions for the process $D_s^+ \to \pi^+\pi^0\pi^0$. The experimental datas of BESIII are represented by points with error bars, and the fitting results of theoretical calculations by the solid black lines. The colored dashed lines indicate the components of the theoretical model.}}
  	 \label{Fig:mass}
 \end{figure}
  
   \begin{figure}[htpb]
	\begin{center}
	\includegraphics[scale=0.8]{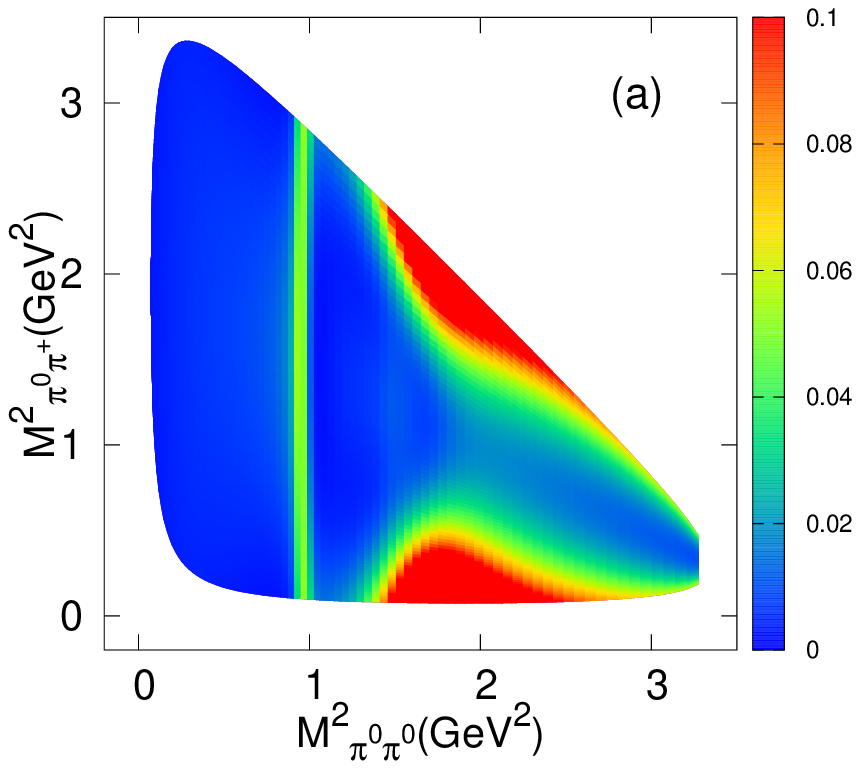}
	\includegraphics[scale=0.8]{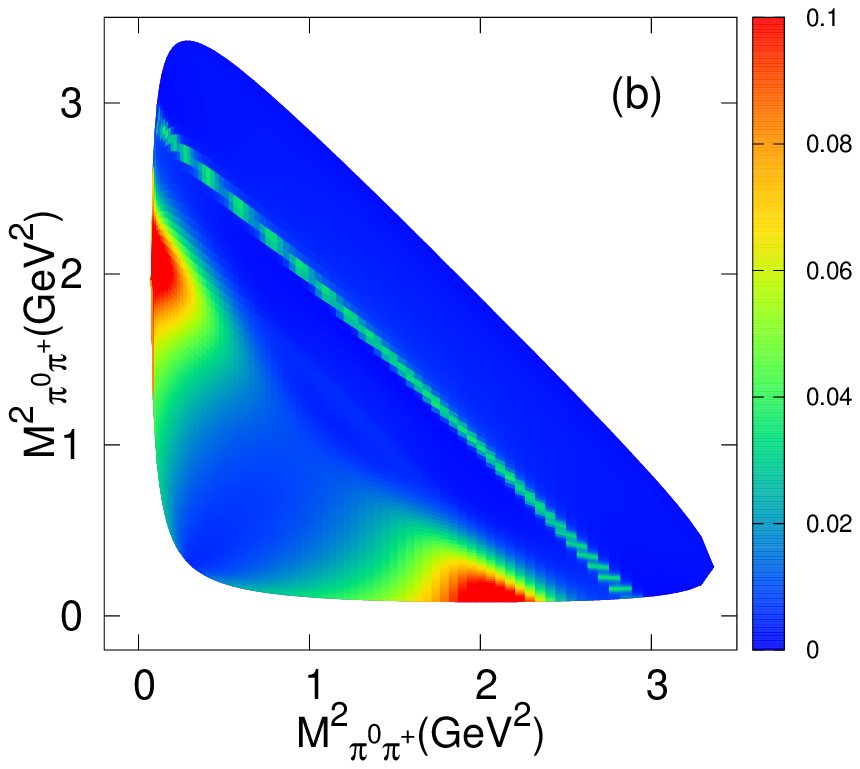}
	\end{center}
	\caption{\small {The Dalitz plots of `$M^2_{\pi^+\pi^0}$' vs.~`$M^2_{\pi^0\pi^0}$' (a) and `$M^2_{\pi^+\pi^0}$' vs.~`$M^2_{\pi^+\pi^0}$' (b) for the process $D^+_s\to \pi^+\pi^0\pi^0$.}}
	\label{fig:dalitz}
\end{figure}
In our model, there are five free parameters, (1) $V_p$, the global normalization of $f_0(980)$ in Eq.~(\ref{eq:fullamp}), (2) the constant $a_{\rm bg}$ as the background contribution, (3) $\alpha$ corresponding to the strength of the $f_0(1370)$ amplitude in Eq.~(\ref{eq:1370}), (4) the width $\Gamma_{f_{0}(1370)}$, (5) $\beta$ as  the strength of the $f_2(1270)$ amplitude in Eq.~(\ref{eq:termD}). In order to present our numerical results, we fit the free parameters to the BESIII measurements of the $\pi^0\pi^0$ and $\pi^+\pi^0$ invariant mass distributions. Here we have $V_p=200$, $a_{\rm bg}=300$, $\Gamma_{f_0(1370)}=236$~MeV, $\alpha=85$, and $\beta=250$.\footnote{The free parameters $V_p$, $a_{\rm bg}$, $\alpha$, and $\beta$ could be complex, which will affect the interferences of the different contributions. Considering our results could reasonably describe the experimental results, we take them to real in order to reduce the numbers of the free parameters.}

Then we have calculated the $\pi^0\pi^0$ and $\pi^+\pi^0$ invariant mass distributions as shown in Fig.~\ref{Fig:mass}.  
The black points with error bars labeled as `BESIII data' are the BESIII data taken from Ref.~\cite{BESIII:2021eru}, and the solid black curves labeled as `Total' are our theoretical results for the total contributions of Eq.~(\ref{eq:M}). In addition, we present the contributions from the $f_0(980)$, $f_0(1370)$, $f_2(1270)$, and the background, which are labeled as `$f_0(980)$', `$f_0(1370)$', `$f_2(1270)$', and `Background', respectively.  
One can find a significant cusp signal around 980~MeV in the $\pi^0\pi^0$ invariant mass distribution, which can be associated with the scalar $f_0(980)$ dynamically generated from the $S$-wave pseudoscalar-pseudoscalar interaction. The enhancement structure around 1300~MeV could be well described, and mainly comes from the contributions of the $f_0(1370)$ and $f_2(1270)$. Indeed, the mass and width of the $f_0(1370)$ have large uncertainties, and it is difficult to extract the exact properties of $f_0(1370)$ due to the overlap with the signal of $f_2(1270)$. 
Thus, the more precise of the $\pi^0\pi^0$ mass distribution of $D^+_s\to \pi^+\pi^0\pi^0$ should be more helpful to shed light on the nature of the resonance $f_0(1370)$ and $f_2(1270)$.

For the $\pi^+\pi^0$ invariant mass distribution, our results are in agreement with the BESIII results considering the large experimental uncertainties. Indeed, one could find a structures around 1.6~GeV, which could be due to resonance $\rho_(1700)$. Since the BESIII measurements have large uncertainties, we do not consider the possible contribution from the resonance $\rho(1700)$ in this work, and one could  perform the complicate calculations when more precise data is available.

We also present the Dalitz plots of `$M^2_{\pi^+\pi^0}$' vs.~`$M^2_{\pi^0\pi^0}$' and `$M^2_{\pi^+\pi^0}$' vs.~`$M^2_{\pi^+\pi^0}$' for the process $D^+_s\to \pi^+\pi^0\pi^0$ in Figs.~\ref{fig:dalitz}(a) and~\ref{fig:dalitz}(b), and one can easily find the contributions of the $f_0(980)$, $f_0(1370)$, and $f_2(1270)$. BESIII has reported the Dalitz plot of `$M^2_{\pi^+\pi^0}$' vs.~`$M^2_{\pi^+\pi^0}$', and one can find our results are consistent with the BESIII measurements.


\section{Conclusions} \label{sec:Conclusions}
Motivated by the recent BESIII measurments about the decay $D_s^+ \to \pi^{+}\pi^{0}\pi^{0}$~\cite{BESIII:2021eru}, we have investigated this Cabibbo-favored process by considering the $S$-wave pseudoscalar-pseudoscalar interactions within the chiral unitary approach, which dynamically generates the scalar $f_0(980)$. In addition, the contributions from the intermediate resonances  $f_0(1370)$ and $f_2(1270)$ are taken into account. 

we have calculated the $\pi^0\pi^0$ and $\pi^+\pi^0$ mass distributions, and find a peak around 980~MeV, which could be associated with the scalar $f_0(980)$, and an enhancement structure around 1300~MeV, mainly due to the intermediate resonances $f_0(1370)$ and $f_2(1270)$. Our results of both the $\pi^0\pi^0$ and $\pi^+\pi^0$ invariant mass distributions are in agreement with the BESIII measurements. In addition, we have predicted the  Dalitz plots of `$M^2_{\pi^+\pi^0}$' vs.~`$M^2_{\pi^0\pi^0}$' and `$M^2_{\pi^+\pi^0}$' vs.~`$M^2_{\pi^+\pi^0}$'.


\begin{acknowledgments}
We warmly thank Prof. Ju-Jun Xie and Prof. Li-Sheng Geng for useful discussions.
This work is supported by the National Natural Science Foundation of China under Grant No. 12192263,  the Natural Science Foundation of Henan under Grand No. 222300420554, the Project of Youth Backbone Teachers of Colleges and Universities of Henan Province (2020GGJS017), the Youth Talent Support Project of Henan (2021HYTP002), and the Open Project of Guangxi Key Laboratory of Nuclear Physics and Nuclear Technology, No.NLK2021-08.
\end{acknowledgments}



\end{document}